\documentclass[superscriptaddress,floatfix,twocolumn,amsmath,amssymb]{revtex4-2}

\usepackage{graphicx}
\usepackage{dcolumn}
\usepackage{bm}
\usepackage{amssymb}
\usepackage{natbib}
\usepackage{amsmath}
\usepackage{mathtools}
\usepackage{color}
\usepackage{datetime}
\usepackage{footnote}
\usepackage{bbold}
\usepackage[normalem]{ulem}
 \usepackage{ragged2e}
 \usepackage{xfrac}
 \usepackage{sidecap}
 \usepackage{setspace}
 \usepackage{multirow}
 \usepackage{xcolor}
 \usepackage{flushend}
\usepackage{hyperref}
\usepackage{cleveref}
\usepackage{microtype}
\usepackage{siunitx}
\usepackage{braket}

\begin{document}

\author{Yang Yang}
\address{Quantum Photonics Laboratory and Centre for Quantum Computation and Communication Technology, RMIT University, Melbourne, VIC 3000, Australia}

\author{Robert J. Chapman}
\address{Quantum Photonics Laboratory and Centre for Quantum Computation and Communication Technology, RMIT University, Melbourne, VIC 3000, Australia}
\address{ETH Zurich, Optical Nanomaterial Group, Institute for Quantum Electronics, Department of Physics, 8093 Zurich, Switzerland}

\author{Akram Youssry}
\address{Quantum Photonics Laboratory and Centre for Quantum Computation and Communication Technology, RMIT University, Melbourne, VIC 3000, Australia}

\author{Ben Haylock}
\address{Centre for Quantum Computation and Communication Technology (Australian Research Council),
Centre for Quantum Dynamics, Griffith University, Brisbane, QLD 4111, Australia}
\address{Institute for Photonics and Quantum Sciences, SUPA,
Heriot-Watt University, Edinburgh EH14 4AS, United Kingdom}

\author{Francesco Lenzini}
\address{Centre for Quantum Computation and Communication Technology (Australian Research Council),
Centre for Quantum Dynamics, Griffith University, Brisbane, QLD 4111, Australia}
\address{Institute of Physics, University of Muenster, 48149 Muenster, Germany}

\author{Mirko Lobino}
\address{Centre for Quantum Computation and Communication Technology (Australian Research Council),
Centre for Quantum Dynamics, Griffith University, Brisbane, QLD 4111, Australia}
\address{Department of Industrial Engineering, University of Trento, via Sommarive 9, 38123 Povo, Trento, Italy}
\address{INFN–TIFPA, Via Sommarive 14, I-38123 Povo, Trento, Italy}

\author{Alberto Peruzzo}
\email[]{alberto.peruzzo@rmit.edu.au}
\address{Quantum Photonics Laboratory and Centre for Quantum Computation and Communication Technology, RMIT University, Melbourne, VIC 3000, Australia}
\address{Qubit Pharmaceuticals, Advanced Research Department, Paris, France}

\title{Programmable quantum circuits in a large-scale photonic waveguide array}

\begin{abstract}
Over the past decade, integrated quantum photonic technologies have shown great potential as a platform for studying quantum phenomena and realizing large-scale quantum information processing. Recently, there have been proposals for utilizing waveguide lattices to implement quantum gates, providing a more compact and robust solution compared to discrete implementation with directional couplers and phase shifters. We report on the first demonstration of precise control of single photon states on an $11\times 11$ continuously-coupled programmable waveguide array. Through electro-optical control, the array is subdivided into decoupled subcircuits and the degree of on-chip quantum interference can be tuned with a maximum visibility of 0.962$\pm$0.013. Furthermore, we show simultaneous control of two subcircuits on a single device. Our results demonstrate the potential of using this technology as a building block for quantum information processing applications.

\end{abstract}

\maketitle


\section{Introduction}
Implementing a universal set of quantum gates is a crucial requirement for a physical system in the standard framework of quantum computing~\cite{DiVincenzo_2000}. Linear optical circuits offer a versatile platform to perform quantum computing tasks~\cite{knill_scheme_2001} and have evolved from free space optics to integrated photonic circuits~\cite{wang_integrated_2020}.
An essential requirement for such technology is the ability to generate controllable and high-visibility quantum interference between single photons with tunable beam splitters~\cite{obrien_photonic_2009}. This can be achieved with Mach-Zehnder interferometers (MZIs) consisting of a phase shifter and two balanced directional couplers (DCs).

In large-scale integrated systems, MZIs equipped with extra phase shifters~\cite{Miller_2015} play a fundamental role in manipulating quantum states~\cite{politi_silica--silicon_2008,matthews_manipulation_2009,Shadbolt:2011bw}.
In addition, large-scale optical networks consisting of several electrically tunable MZIs and phase shifters -namely, a Reck~\cite{reck_experimental_1994} or Clements scheme~\cite{clements_optimal_2016} - are required for realizing quantum photonic processors able to perform any arbitrary linear optics operations on quantum states of light~\cite{carolan_universal_2015,bao2023very}. 
Beyond quantum computing, programmable photonic circuits~\cite{bogaerts_programmable_2020_nature} are a key component in numerous other emerging technologies, including quantum transport simulations~\cite{harris_quantum_2017}, microwave photonics~\cite{marpaung_integrated_2019} and optical neural networks~\cite{shen_deep_2017}.


The schemes based on MZIs suffer from bend losses proportional to the depth of the circuit, and they are not robust against fabrication errors~\cite{saygin_robust_2020,skryabin_waveguide-lattice-based_2021,petrovic_new_2021,tanomura_scalable_2022}. Furthermore, the bending radii are typically significantly longer than the coupling length~\cite{politi_silica--silicon_2008}, meaning the circuit area is dominated by routing waveguides that do not contribute to the logical operation. Photonic waveguide arrays (WA) have been proposed as an alternative that overcomes the aforementioned challenges. WA are periodic structures composed of optical waveguides evanescently coupled to each other~\cite{christodoulides_discretizing_2003}. Since the first demonstration of a two-photon continuous quantum walk on a WA with 21 waveguides~\cite{peruzzo_quantum_2010}, there have been numerous advancements in the field of quantum applications using WAs, ranging from topologically protected quantum state generation~\cite{PhysRevA.92.033815, PhysRevA.105.023513, blanco2018topological} to quantum state processing~\cite{chapman_experimental_2016,tambasco_quantum_nodate, Blanco_Redondo_2020_topological_q_curcuits}. Additionally, theoretical and experimental work has explored the feasibility of using waveguide arrays for implementing 1 and 2-qubit gates~\cite{compagno_toolbox_2015, lahini_quantum_2018, chapman2023quantum}. These structures can also simulate a wide range of condensed matter physics effects~\cite{PhysRevLett.83.4756,PASPALAKIS200630, PhysRevLett.101.193901, lahini_anderson_2008, lahini_observation_2009} as well as offer the possibility of directly implementing tri-diagonal Hamiltonians, in a direct (analog) instead of digital way. Recently, we have demonstrated a reconfigurable waveguide array based on the lithium niobate photonics platform, implementing multiple condensed matter physics models~\cite{yang_topological_2023}. 

In this paper, we demonstrate the implementation of reconfigurable single-qubit gates on an 11-waveguide electro-optically reconfigurable waveguide array (11-RWA), by decoupling pairs of waveguides and implementing subcircuits in the form of tunable DCs within the array. Additionally, we tune the degree of quantum interference with a maximum visibility of 0.962$\pm$0.013 in the subcircuit. Finally, we demonstrate the possibility of using the chip to decouple multiple pairs of waveguides to control multiple single qubits in parallel. Our technology enables large-scale photonic quantum information processing based on reconfigurable continuously-coupled waveguide arrays, providing a platform with improved fabrication error tolerance and reduced bending-loss ~\cite{saygin_robust_2020,petrovic_new_2021,tanomura_scalable_2022}, and direct access to and control of the Hamiltonian terms and evolution~\cite{yang_topological_2023}.

\begin{figure}
    \centering
    \includegraphics[width=1\columnwidth]{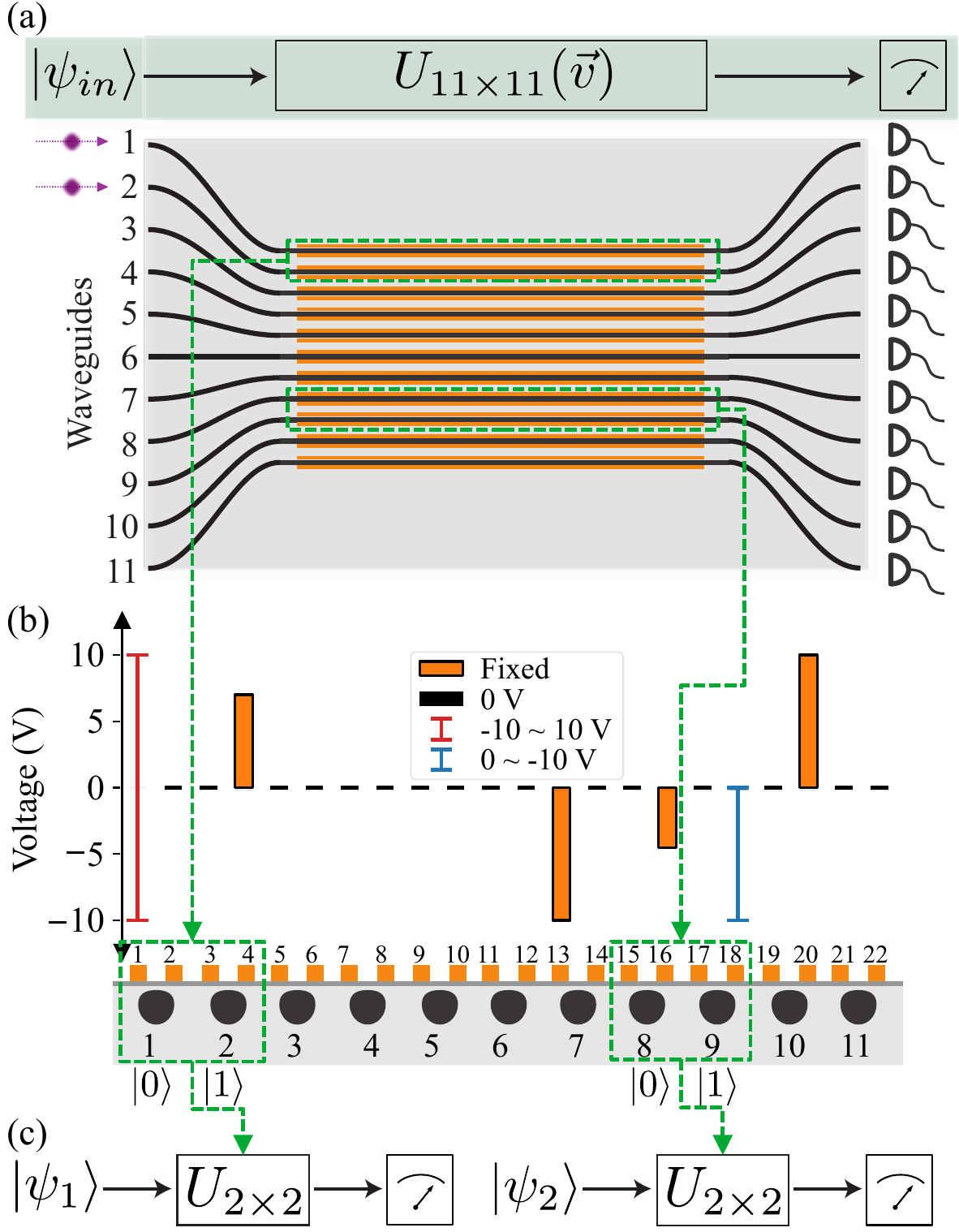}
    \caption{\textbf{Schematic of the 11-RWA and protocol for independent control of subcircuits DC$_{1,2}$ and DC$_{8,9}$.} (a) The 11-RWA has 11 waveguides (black), which implements a voltage-dependent unitary $U_{11{\times}11}(\vec{v})$,  programmed via electric fields applied to the electrodes (orange). (b) Electrode configuration to implement the experiments in this work, the cross-section of the RWA, and the schematic of subcircuits DC$_{1,2}$ and DC$_{8,9}$. By decoupling optical modes 1 and 2 from the other modes of the circuit, WG$_1$ and WG$_2$ act as an independent coupled system DC$_{1,2}$. Path encoding is used in this work, i.e., $|0\rangle=\begin{bmatrix}1&0\end{bmatrix}^T$ is encoded as a photon in optical mode 1, $|1\rangle=\begin{bmatrix}0&1\end{bmatrix}^T$ is encoded as a photon in optical mode 2. (c) By decoupling and post-selection, the single photon input states $\ket{\psi_1}$ and $\ket{\psi_2}$ are processed independently by two $2\times 2$ unitaries, effectively implementing two single-qubit quantum gate operations.}
    \label{fig:chip_schematic}
\end{figure}

\section{results}

\subsection{Reconfigurable Waveguide arrays via electro-optic control} 
The 11-RWA used in this work is fabricated on x-cut lithium niobate, with a designed continuously-coupled region of 24~mm. The reconfigurability of the device is enabled by creating electric fields across the cross-section of the array via electrodes positioned on the top of the waveguides. Further details about the device can be found in the supplementary materials.

An ideal 11-RWA device shown in Fig~\ref{fig:chip_schematic}(a) can be modeled by the voltage-dependent tri-diagonal Hamiltonian \cite{christodoulides_discretizing_2003}
\begin{equation}
H_{11{\times}11}(\vec{v}) = \begin{bmatrix} 
    \beta_{1} & C_{1,2} & 0 &  \dots & 0 & 0\\
    C_{1,2} & \beta_{2} & C_{2,3} &  \dots & 0 & 0\\
    0 & C_{2,3} & \beta_{3} & \dots & 0 & 0\\
    \vdots & \vdots & \vdots & \ddots & \vdots &\vdots\\
    0 & 0 & 0  & \dots & \beta_{10} & C_{10,11} \\
    0 & 0 & 0 & \dots & C_{10,11} & \beta_{11} \\
    \end{bmatrix},
    \label{1.1.1}
\end{equation}
where the propagation constant $\beta_n$ and coupling constants C$_{n,n+1}$ between adjacent waveguides (WG) ($n$ indicate the waveguide label) can be controlled via electric field generated by $\vec{v}=\left(V_1, V_2,..V_{22}\right)$ from 22 electrodes, which changes the refractive index of the material via the Pockels effect.
The unitary transformation of the device is given by
\begin{align}
    U_{11{\times}11}(\vec{v})=e^{-iH_{11{\times}11}(\vec{v})L},
    \label{1.1.2}
\end{align}
where L is the effective coupling length of the device.

We used a fiber-coupled polarized 808~nm laser diode for classical characterization of the device and the setup used to control the device is illustrated in the supplementary materials. The voltage corresponds to the amplitude of the non-biased square pulse. Further details can be found in the supplementary materials.

In practice, this model does not apply accurately to a real device due to fabrication imperfections~\cite{youssry_experimental_2022}. To control the device, a model-based machine learning approach~\cite{youssry_experimental_2022}, or model-free approaches can be used. In this paper, we obtain the controlling voltages by using a model-free approach in which measurement-based lookup maps are built for different configurations and then those maps are used to search for the required control voltages satisfying given criteria (such as a target reflectivity with minimum leakage). Details can be found in the supplementary materials.

\begin{figure}[h!]
    \centering
    \includegraphics[width=1\columnwidth]{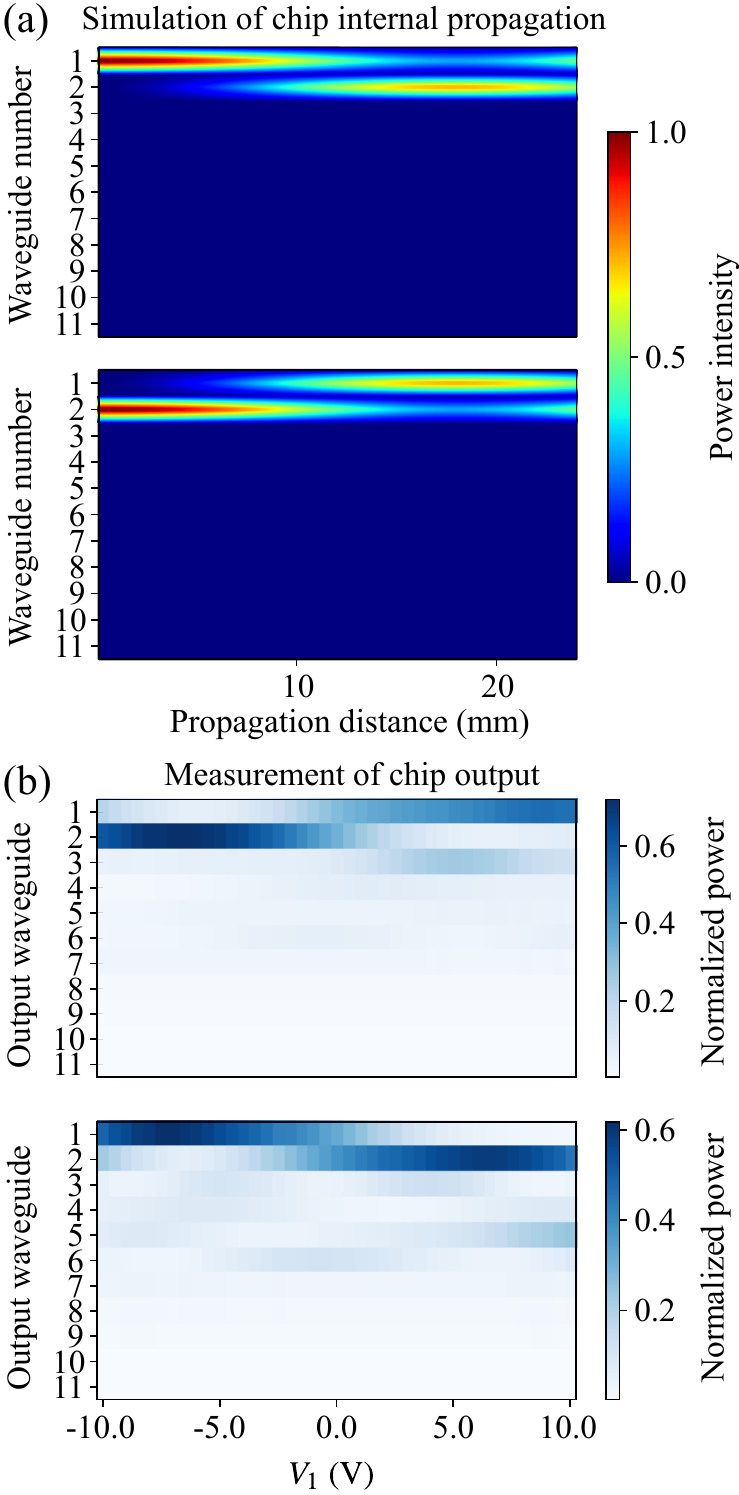}
    \caption{\textbf{Decoupling and control of subcircuit DC$_{1,2}$.} (a) Numerical simulations of the propagation of light inside the chip for input waveguides 1 (top) and 2 (bottom), where DC$_{1,2}$ is decoupled and $V_1$=0~V. (b) Experimentally measured optical output power of the RWA for input waveguides 1 (top) and 2 (bottom), as a function of the voltage on electrode 1 based on the configuration for in Fig~\ref{fig:chip_schematic} (c).}
    \label{fig:dc12_results}
\end{figure}

\subsection{Subcircuit control in an RWA} 
One of the key requirements for implementing universal multiport interferometers \cite{reck_experimental_1994,clements_optimal_2016} using RWAs, is the capability of splitting an RWA into decoupled subcircuits as shown in Fig~\ref{fig:chip_schematic} (i.e. decoupling a subset of the waveguides from the rest, and being able to control that subset independently), as well as implementing arbitrary $2{\times}2$ unitary operations.

To decouple a subcircuit, the coupling constants between the boundary waveguides of the target subcircuit and the rest of the RWA need to be reduced to zero. This will result in a structure of three subcircuits with the Hamiltonian reduced to a block diagonal matrix of the form $H_{11\times 11} = H_1 \oplus H_2 \oplus H_3$. Consequently, the resulting unitary will also be a block diagonal of the form $U_{11\times 11} = U_1 \oplus U_2 \oplus U_3$. Here, we show the decoupling of a $2\times 2$ subcircuit which provides the basis for qubit operations as well as the main building block of higher-dimensional unitary decomposition methods. A general $2{\times}2$ unitary can  be represented as: 
\begin{align}
R_zU_{DC}=\begin{bmatrix}1 & 0\\0 & e^{i\phi}\end{bmatrix}\begin{bmatrix}\sqrt{\eta} & i\sqrt{1-\eta}\\i\sqrt{1-\eta} & \sqrt{\eta}\end{bmatrix},
\label{1.1.3}
\end{align}
where U$_{DC}$ is the unitary of a tunable DC with $\eta$ indicating its reflectivity and R$_{z}$ is the unitary transformation of a phase shifter with $\phi$ defining the phase. In our technology, both $\eta$ and $\phi$ are voltage-dependent, and, in this work, we restrict our measurements to $\eta$.

In practice, achieving a coupling constant of zero might be difficult with a limited voltage range. This results in imperfect decoupling and the leakage of optical power from a target subcircuit to the rest of the waveguides. The leakage can be quantified using the definition:
\begin{equation}
    P_{\text{leakage}} (\%) = \sum_{i=1,i{\neq}k,k+1}^{N}P_i (\%),
\end{equation}
where $P_i$ is the percentage normalized measured output power at WG$_i$, the index $k$ is the subcircuit index of interest and $N$ is the total number of waveguides. 
In this paper, we demonstrate the decoupling of the subcircuit consisting of WG$_1$ and WG$_2$(as shown in Fig~\ref{fig:chip_schematic}). Since this is an edge subcircuit, the RWA will be split into two subcircuits rather than three. We empirically decreased the value of $C_{2,3}$ in Eq\ref{1.1.1} 
by applying a voltage of 7~V to electrode 4 ($V_4$=7~V) on the RWA. Specifically, for light input in WG$_1$, the leakage is reduced from 63\% to 33\%, and for light input in WG$_2$, it is reduced from 81\% to 35\%, compared to the scenario where 0 is applied to electrode 4 ($V_4$=0~V, see Fig~\ref{fig:chip_schematic}(b) and supplementary materials). We varied the propagation constant $\beta_1$ of WG$_1$ by sweeping the $V_1$ from -10 to 10~V (see Fig~\ref{fig:chip_schematic}(b) and supplementary materials) to control the reflectivity $\eta$ of DC$_{1,2}$, to obtained to realize a subcircuit that implements a tunable DC within the 11-RWA. 

Fig~\ref{fig:dc12_results}(a) shows the ideal simulations of the light propagation for light launched into WG$_1$ and WG$_2$ when only the decoupling voltage (on electrode 4) is applied. This simulation matches the measured reflectivity in Fig~\ref{fig:dc12_results}(b), but does not restrict the decoupling voltage amplitude. With experimentally restricted decoupling voltage, part of the light can leak to the rest of the RWA during the propagation. The simulated relationship between leakage and decoupling voltage can be found in supplementary materials.
In Fig~\ref{fig:dc12_results}(b), we report the experimental measurements of the voltage-dependent output power from the 11-RWA with light input in WG$_1$ and WG$_2$ respectively.

\begin{figure}[h!]
    \centering
    \includegraphics[width=1\columnwidth]{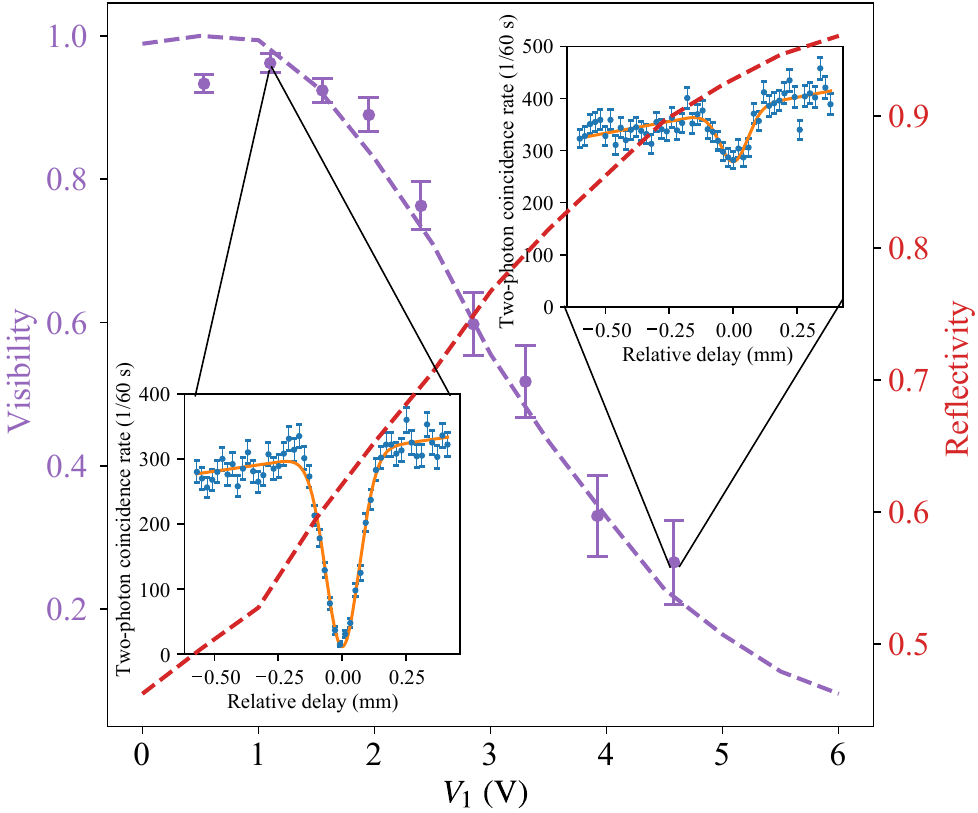}
    \caption{\textbf{The Hong-Ou-Mandel (HOM) interference measurements as a function of $V_1$.} 
    The insets show two-photon interference experiments, performed at two different voltages that are applied to electrode 1, where the blue points represent the measured two-photon coincidence rate integrated over 60~s as a function of the relative delay, and the orange line is the fitting. The error bars represent the standard deviation based on the assumption of Poissonian photon statistics. The left inset shows a high-visibility configuration, while the right inset shows a low-visibility one. The visibility is estimated from the fitting and plotted as a function of the voltage shown as the purple points. The error bars of the visibility are based on the calculations detailed in the Supplementary Materials. The red line is the measured reflectivity as a function of voltage of this configuration based on Fig~\ref{fig:dc12_results}(b). The purple dashed line shows the theoretical visibility versus voltage calculated from the reflectivity at the same voltage. }
    \label{fig:hom_dip}
\end{figure}

\subsection{Reconfigurable quantum interference} 
The capability of generating high-visibility two-photon quantum interference is crucial for photonic quantum technologies~\cite {doi:10.1126/science.1142892,10.1116/5.0007577}.
We performed reconfigurable quantum interference~\cite{hong_measurement_1987} experiments within the subcircuit DC$_{1,2}$, providing a proof-of-concept for continuously-coupled integrated photonic systems. Photon pairs at wavelength 807.5~nm generated by a spontaneous parametric down-conversion source (see supplementary materials) are launched into WG$_1$ and WG$_2$ with the input state $\frac{1}{\sqrt{2}}(|01\rangle+|10\rangle)$ as shown in Fig~\ref{fig:chip_schematic}(a) (encoding scheme is explained in the Fig~\ref{fig:chip_schematic} caption). We measured the two-photon coincidence rate while scanning a physical delay line of one of the photon beams at different reflectivity of DC$_{1,2}$. Ideally, when $\eta$=0.5, two photons come out from the same waveguide with the output state $\frac{1}{\sqrt{2}}(|00\rangle+e^{i\psi}|11\rangle)$ at a unity probability due to `photon bunching' effect~\cite{Beugnon_2006}, which gives zero coincidence counts between two waveguide outputs.

The visibility is a measure of the contrast of the HOM `dip' in quantum interference experiments. Each experiment is done by measuring the two-photon coincidence counts as a function of relative delay between two-photon paths at a fixed voltage configuration as shown in the insets in Fig~\ref{fig:hom_dip} (blue data points). A Gaussian function with a linear term is used to fit the measurement results to extract the visibility~\cite{laing_high-fidelity_2010} (shown as the orange lines). Details about the fitting procedure are provided in the supplementary materials. The insets show two examples of such an experiment at two different voltages corresponding to two different reflectivities. The visibility at reflectivity $\eta=0.496$ ($V_1$=0.5~V) is $\bar{V}=0.962\pm0.013$. The visibility at reflectivity $\eta=0.897$ ($V_1$=4.5~V) is $\bar{V}=0.265\pm0.058$. The reflectivity $\eta$ as a function of voltage is calculated from the power measurements in Fig~\ref{fig:dc12_results}(b) and is shown as the red dashed line. The procedure is then repeated to sweep the visibility measurement (shown as the purple data points) over a voltage range that corresponds to the reflectivities ranging from 0.5 to 1. Finally, the ideal visibility as a function of voltage is calculated from the reflectivity at the same voltage and shown as the purple dashed line. The calculations can be found in the supplementary materials. The results show that the measured visibility is consistent with the ideal calculated visibility.

\begin{figure*}[h!]
    \centering
    \includegraphics[width=1.8\columnwidth]{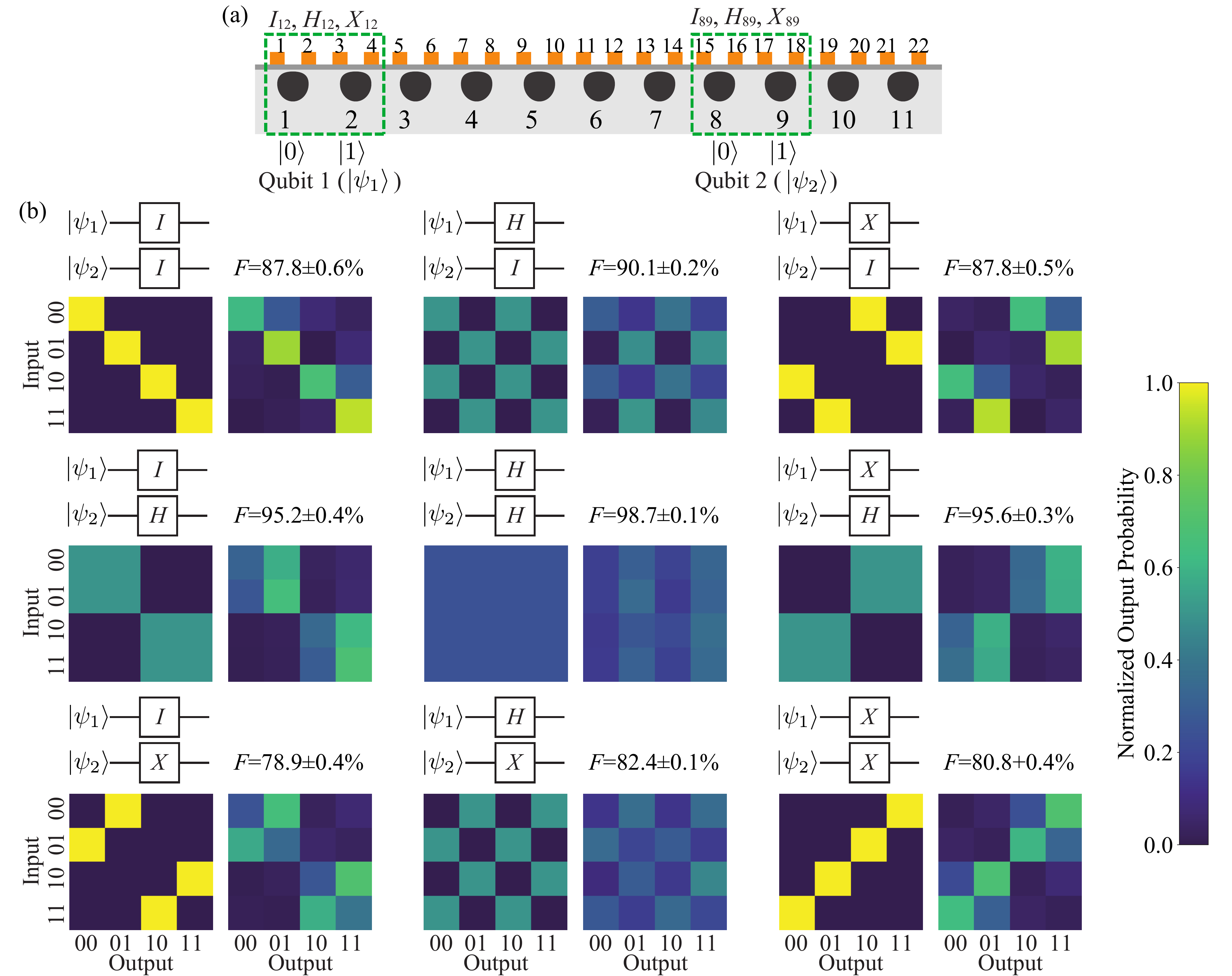}
    \caption{\textbf{Parallel operation of two single-qubit gates in waveguide array subcircuits.} 
    (a) Schematic of the experiment implementing two tunable subcircuits. By selecting the reflectivity, single qubit gates such as the identity $I$, Hadamard $H$, or Pauli $X$ are realized for two path-encoded qubits $\ket{\psi_1}$ and $\ket{\psi_2}$.
    (b) Single-qubit gates ($I$, $H$, and $X$ gate) are implemented by controlling DC$_{1,2}$ and DC$_{8,9}$, with two balanced laser inputs in parallel. For each operation, we show the ideal map (left) and the experimentally measured map (right). We also report the post-selected average fidelity $F$ over all input states.}
    \label{fig:parallel}
\end{figure*}

\subsection{Parallel quantum gates} 
Controlling multi-qubit systems is crucial to advanced quantum technologies~\cite{jianwei_multidimentional_2018}.
We demonstrate this capability by running single-qubit operations on two independent qubits simultaneously~\cite{Shadbolt:2011bw} within the RWA. We implement a second subcircuit DC$_{8,9}$ as shown in Fig~\ref{fig:chip_schematic}(c) and Fig~\ref{fig:parallel}(a) with the electrode configuration shown in Fig~\ref{fig:chip_schematic}(b). Decoupling voltages are applied on $V_{13}$ and $V_{20}$ and $V_{16}$=-4.5~V is used to shift the reflectivity curve to gain a larger tunable reflectivity range.

To showcase the capability of implementing the single-qubit gates in parallel on DC$_{1,2}$ and DC$_{8,9}$, we launched two balanced laser beams into two subcircuits separately by splitting the laser into two beams with a 50/50 808~nm fiber-coupled beam splitter and measure the output power for 4 different input combinations. We built a lookup map by varying both tuning electrodes at the same time (see supplementary material).
The reflectivity required for implementing single-qubit $X$, $H$, and $I$ gates are $\eta=$ 0, 0.5, and 1. We found the corresponding voltages based on linear fitting of the reflectivity curves of each DC based on the measured lookup map to run parallel quantum gates. In Fig~\ref{fig:parallel}(b), we report the ideal and measured truth tables and the fidelity for all combinations of two single-qubit gates based on post-selected results by ignoring the leakage.

The post-selected average fidelity (over input states $i\in \{0,1,2,3\} \equiv \{ \ket{00}, \ket{01}, \ket{10}, \ket{11}\} $) for the parallel operation is given by $F = \frac{1}{4}\sum_{i=1}^4 F_i$, where $F_i = \sum_{j=1}^{4} \sqrt{P_{i,j}^{T} P_{i,j}^{M}} $ is the fidelity between the target and measured output power distributions for the input state $i$, and the summation is over the output states $j\in \{0,1,2,3\} \equiv \{ \ket{00}, \ket{01}, \ket{10}, \ket{11}\}$. The normalized power distribution is calculated as $P_{i,j}^M = \frac{M_{I(j)}}{\sum_{k\in{1,2,8,9}} M_k}$, where $I(j):\{0,1,2,3\}\to \{1,2,8,9\}$ is the index mapping between state encoding $j$ and the waveguide index, and $M_k$ is the measured power at output waveguide $k$.

The post-selected average fidelity of all parallel quantum gates experiments is 88.5\%. They are relatively good except in some cases when $I$ and $X$ gates on DC$_{8,9}$ are involved. 
This is because the lookup maps, created based on limited control, do not fully cover the space of the unitaries that can be implemented by the device and lack the solutions for achieving high-fidelity $I$ and $X$ gates on DC$_{8,9}$. These results demonstrate the proof of concept, but further optimization is necessary to find better control voltages.

\section{Discussion} 
In this paper, we have demonstrated reconfigurable quantum interference and parallel single-qubit operations on a large-scale RWA.
Large-scale WAs were static~\cite{lahini_anderson_2008,lahini_observation_2009} in the past or not in a fully controllable fashion~\cite{hoch_reconfigurable_2022}. Utilizing thermal-optic effects is currently the prevailing method for constructing reconfigurable photonic circuits. However, it cannot be employed to construct a fully controllable waveguide array device due to the diffusive propagation of the thermo-optic effect along the material~\cite{https://doi.org/10.1002/lpor.202000024}. In contrast, the electric field is confined within the material due to the shielding effect from neighboring electrodes, enabling stable and precise control in electro-optic devices~\cite{10011218} without the need for thermalization.

To expand the device's capabilities for quantum computing applications, implementing a two-qubit CNOT gate is essential. There are several potential pathways to achieve this. One approach is to directly locate the target unitary of a two-qubit gate in lookup maps by spanning its controlling parameters or using other machine learning methods~\cite{youssry_modeling_2020,lazin2023high,youssry_experimental_2022}. Another option is to design a device that naturally implements the CNOT gate without any applied control~\cite{lahini_quantum_2018,chapman2023quantum}. However, this approach faces controllability issues when adding tunability. Additionally, the CNOT gate can be decomposed into three sections, with each section implementing single-qubit gates exclusively~\cite{o2003demonstration}. This can be achieved by cascading three sections of RWAs, with each RWA needing to implement single-qubit gates.

To further improve the performance and scalability of such devices, efforts need to be put into device design and fabrication as well. For example, the newly developed z-cut thin film lithium niobate platform~\cite{zhang_integrated_2021} allows different electrode patterns and reduced footprint while offering the capability of on-chip single-photon sources~\cite{White:19,10.1063/1.5054865} and detectors~\cite{Lomonte:2021}. It is worth noting that WAs have been extensively used in demonstrating topological effects. These effects can be leveraged to generate topologically protected quantum states and build topologically robust quantum gates~\cite{rechtsman2016topological,tambasco_quantum_nodate,blanco2018topological,wang2019topological_2}, offering an alternative approach to implementing universal gate sets.


The measurement-based lookup map method we used in this work addresses the challenge of the requirement of having an accurate mathematical model for the fabricated device. In this paper, we focused on building maps for reflectivity and leakage using power measurements. The method can be extended to settings where the phase information is also available. This will allow better characterizations of the leakage and light power crosstalk between subcircuits. The phase information can be acquired with either classical light only~\cite{rahimi2013direct,youssry_modeling_2020,hoch2023characterization} or quantum light via HOM measurements~\cite{peruzzo2011multimode,laing2012super,dhand2016accurate}. In this work, we fixed the voltage steps applied to two electrodes, creating a uniform resolution over the range of the map. To have better control over the chip and reduce the leakage or widen the reflectivity range, we can use more electrodes and a larger voltage range. In this case, to overcome the computational challenge of building the lookup tables, we can use non-uniform/adaptive sampling to choose the resolution efficiently. Additionally, there are machine learning methods (such as Bayesian optimization \cite{lazin2023high}) that automate the process of building the lookup maps and the optimization simultaneously.

Another approach that should be considered is cascading more sections of the chip. It can be shown theoretically, that adding more sections can result in better gate fidelity based on quantum controllability arguments. The post-selected fidelity can decrease further when cascading multiple sections and increasing the number of operating qubits. In such cases, it is crucial to consider and compensate for the leakage from the previous sections in the subsequent sections with advanced control methods.
Thus, the results we presented here show the potential of such a device as a fundamental building block toward arbitrary unitary~\cite{saygin_robust_2020,tanomura_scalable_2022} or an alternative way to implement the discrete scheme~\cite{reck_experimental_1994,clements_optimal_2016}, providing the advantages of reduced bending-loss (see further discussion in the supplementary materials) and fabrication error robustness.

Other potential applications besides the ones presented in this paper include high-dimensional quantum computing~\cite{10.3389/fphy.2020.589504}, and topological quantum computation~\cite{RevModPhys.80.1083}. Our results point
towards building integrated linear optical quantum circuits in a continuously-coupled way of manipulating photon states with potential applications in quantum metrology, quantum simulations~\cite{aspuru-guzik_photonic_2012}, and quantum information processing.

In summary, this work is the first proof-of-principle demonstration of using continuosly-coupled waveguide arrays, instead of networks of beamsplitters (Reck or Clements scheme), as a new architecture for photonic quantum computing. The presented device enables controllable operations on subcircuits through the application of electrode voltage for channel decoupling. Our achievement of a high-visibility (96\%) two-photon quantum interference within a subarray is a crucial step towards implementing quantum circuits with this technology. Additionally, we successfully demonstrated parallel control of single-qubit gates on two subcircuits. This work establishes a substantial connection between quantum information science and photonic engineering, showcasing the potential of precise control of RWAs for future quantum technologies. These results open up new avenues for further research in the design, fabrication, and control of such devices.

\section*{Acknowledgements}
AP acknowledges an RMIT University Vice-Chancellor’s Senior Research Fellowship and a Google Faculty Research Award. ML was supported by the Australian Research Council (ARC) Future Fellowship (FT180100055). BH was supported by the Griffith University Postdoctoral Fellowship. This work was supported by the Australian Government through the Australian Research Council under the Centre of Excellence scheme (No: CE170100012), and the Griffith University Research Infrastructure Program. This work was partly performed at the Queensland node of the Australian National Fabrication Facility, a company established under the National Collaborative Research Infrastructure Strategy to provide nano- and microfabrication facilities for Australia’s researchers.

\clearpage
\setcounter{figure}{0}
\makeatletter 
\renewcommand{\thefigure}{S\@arabic\c@figure}
\makeatother

\setcounter{equation}{0}
\makeatletter 
\renewcommand{\theequation}{S\@arabic\c@equation}
\makeatother

\setcounter{table}{0}
\makeatletter 
\renewcommand{\thetable}{S\@arabic\c@table}
\makeatother

\end{document}


\title{Supplementary Materials for \\Programmable quantum circuits in a large-scale photonic waveguide array}

\author{Yang Yang}
\address{Quantum Photonics Laboratory and Centre for Quantum Computation and Communication Technology, RMIT University, Melbourne, VIC 3000, Australia}

\author{Robert J. Chapman}
\address{Quantum Photonics Laboratory and Centre for Quantum Computation and Communication Technology, RMIT University, Melbourne, VIC 3000, Australia}
\address{ETH Zurich, Optical Nanomaterial Group, Institute for Quantum Electronics, Department of Physics, 8093 Zurich, Switzerland}

\author{Akram Youssry}
\address{Quantum Photonics Laboratory and Centre for Quantum Computation and Communication Technology, RMIT University, Melbourne, VIC 3000, Australia}

\author{Ben Haylock}
\address{Centre for Quantum Computation and Communication Technology (Australian Research Council),
Centre for Quantum Dynamics, Griffith University, Brisbane, QLD 4111, Australia}
\address{Institute for Photonics and Quantum Sciences, SUPA,
Heriot-Watt University, Edinburgh EH14 4AS, United Kingdom}

\author{Francesco Lenzini}
\address{Centre for Quantum Computation and Communication Technology (Australian Research Council),
Centre for Quantum Dynamics, Griffith University, Brisbane, QLD 4111, Australia}
\address{Institute of Physics, University of Muenster, 48149 Muenster, Germany}

\author{Mirko Lobino}
\address{Centre for Quantum Computation and Communication Technology (Australian Research Council),
Centre for Quantum Dynamics, Griffith University, Brisbane, QLD 4111, Australia}
\address{Department of Industrial Engineering, University of Trento, via Sommarive 9, 38123 Povo, Trento, Italy}
\address{INFN–TIFPA, Via Sommarive 14, I-38123 Povo, Trento, Italy}

\author{Alberto Peruzzo}
\address{Quantum Photonics Laboratory and Centre for Quantum Computation and Communication Technology, RMIT University, Melbourne, VIC 3000, Australia}
\address{Qubit Pharmaceuticals, Advanced Research Department, Paris, France}

\maketitle

\subsection*{Experimental setup}

\begin{figure*}
    \centering
    \includegraphics[width=1.8\columnwidth]{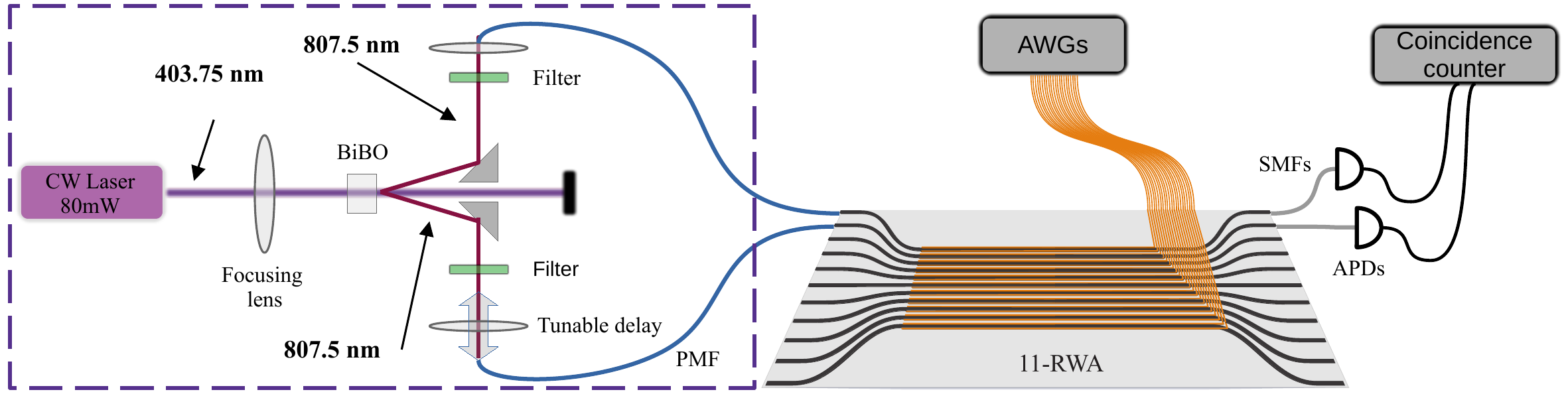}
    \caption{\textbf{Schematic of the quantum interference setup.} A spontaneous parametric down-conversion (SPDC) single photon source (in purple dashed frame) is used to prepare single photon states. A multi-channel AWG output is connected to the electrodes of the 11-RWA via a ribbon cable. Avalanche photodiodes used to detect photons are connected to a coincidence counter.}
    \label{fig:setup}
\end{figure*}

A polarized 808~nm laser and multi-channel fiber-coupled high-speed optical power meter with oscilloscopes are used for classical characterizations. A multi-channel AWG is used to generate pulses that are applied to electrodes to program the 11-RWA. The power distribution of the static classical characterization of the RWA for each input waveguide is extracted from the power intensity measured by a CCD camera at the output facet of the RWA. In the parallel gate measurements, the 808~nm laser is split into two beams via a 50/50 808~nm fiber-coupled beam splitter. 

A schematic of the quantum interference measurements setup is shown in Fig~\ref{fig:setup}. A type 1 free space spontaneous parametric down-conversion (SPDC) single photon source (purple dashed frame) and single photon counting modules (SPCMs) are used to generate and detect photon pairs. The SPCMs are from Excelitas Technologies (SPCM-800-12-FC) with a measured dark count rate $\sim$30/s. Each channel has a counting rate $\sim$8000/s with an expected accidental rate of less than 1 per minute. The transmission loss is primarily attributed to fiber-to-chip coupling, resulting from a mismatch between fiber and waveguide modes. A total transmission loss, which includes propagation loss, was measured at 9.8 dB (90\%). A 1mm BiBO crystal is pumped by a 403.75~nm, 80~mW continuous laser. Two generated photons with an opening angle of 3$^{\circ}$ are separated by two prism mirrors, pass through a narrow band filter with a Full width at half maximum (FWHM) of 3.1~nm and are then coupled into two PMFs. One of the polarization-maintaining fibers (PMFs) is mounted on the motor-controlled stage, which enables a tunable delay. Photon polarization is rotated from horizontal to vertical before photons are coupled into the PMFs. Single-mode fibers (SMFs) are used to couple the chip outputs. The integration time of single-photon experiments is 60~s.

The 11-RWA is fabricated using annealed and reverse proton exchange technology with x-cut bulk lithium niobate. The designed waveguide mode has a diameter of 5~$\mu$m and is separated by a distance of 5~$\mu$m. The gold electrode has a width of 2.5~$\mu$m and is separated by a distance of 2.5~$\mu$m~\cite{yang_topological_2023}. A 200 nm silicon dioxide buffer layer is placed between the lithium niobate and gold electrodes. Further design and fabrication methods can be found in~\cite{Lenzini:15, Lenzini_multiplexing,youssry_experimental_2022}.



Applying controlling voltages on electrodes causes DC drift of LiNbO$_3$ due to electric charge accumulation, which results in a drift of optical output states of the chip~\cite{yamada_dc_1981}.
100~Hz modulated non-biased square pulses generated by AWGs are used to stabilize the optical output states~\cite{lenzini_integrated_2018}.
Each electrode is connected to a channel of arbitrary waveform generators (AWGs) independently. An external trigger is used to synchronize multiple AWGs. 
Voltage-dependent output power distributions in classical characterizations are taken from an average of 20 periods post-selected parts of the square pulse. In quantum interference measurements, the square pulses are continuously applied until a sufficient integration of two-photon coincidence counts is achieved. Long-term drift over tens of hours was measured to be negligible by continuously applying this control method and measuring the chip output states. Additionally, classical light characterization was conducted on different days from the HOM experiments, confirming the repeatability of the experiments.

\begin{figure}
    \centering
    \includegraphics[width=1\columnwidth]{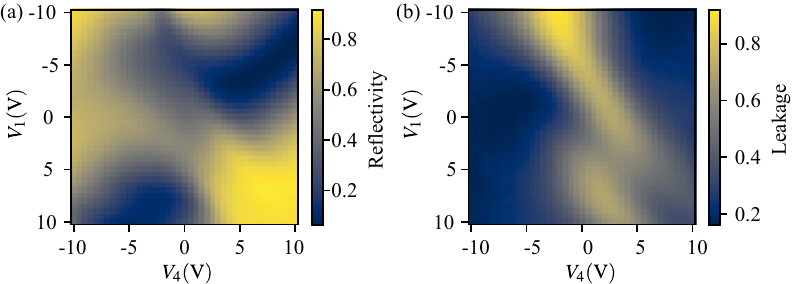}
    \caption{\textbf{Lookup map of subcircuit DC$_{12}$ with input WG$_1$ based on two tunable electrodes 1 and 4, and all others grounded.} (a) Reflectivity map. (b) Leakage map.}
    \label{fig:lookup_map}
\end{figure}

\begin{figure}
    \centering
    \includegraphics[width=1\columnwidth]{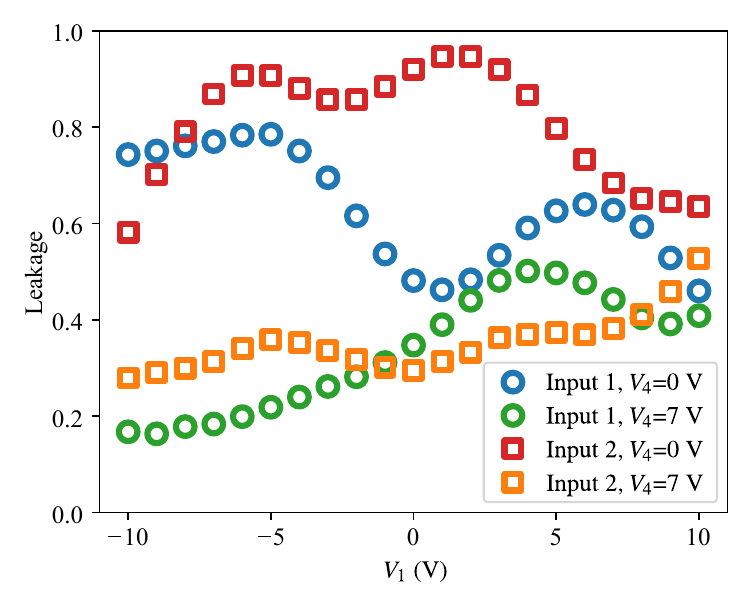}
    \caption{\textbf{The leakage comparison for DC$_{1,2}$.} The leakage is plotted as a function of the voltage applied to electrode 1 for input WG$_1$ and WG$_2$ at decoupling ($V_4=$7~V) and non-decoupling ($V_4=$0~V) conditions. The overall averaged leakage is reduced from 63\% to 33\% for light input in WG$_1$ and from 81\% to 35\% for light input in WG$_2$ by applying 7~V to electrode 4.}
    \label{fig:leakage_analysis}
\end{figure}

\begin{figure}
    \centering
    \includegraphics[width=1\columnwidth]{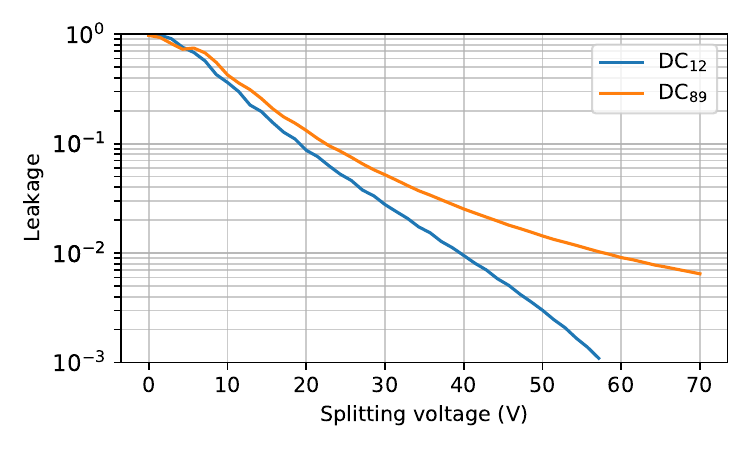}
    \caption{\textbf{Numerical simulation of the leakage of two subcircuits.} With limited voltage in experiments, we expect leakage to be around 40\% for both DCs. We expect less than 1\% leakage when the decoupling voltage is higher than 40~V (DC$_{12}$, input port 1) and 57~V (DC$_{89}$, input port 8) based on the assumption that the coupling constant is linearly-dependent on voltage.}
    \label{fig:leakage}
\end{figure}

\subsection*{Chip decoupling}
Chip decoupling can be achieved by decreasing the coupling strength between boundary waveguides. In practice, the voltage-Hamiltonian dependence deviates from the standard models that are commonly used in this technology. Such models assume a linear dependence, with the Hamiltonian taking a real-valued tri-diagonal form. In order to avoid using an inaccurate model for chip control, we propose another model-free method based on experimental measurements. In particular, we built lookup maps for the reflectivity and leakage as a function of two control voltages (applied to two tunable electrodes). The lookup maps are extracted from laser-based output power measurements with varying voltage and input configurations. Each measurement is corrected using static chip characterization to compensate for imperfect chip-to-fiber coupling on the output side. These maps capture the voltage-dependent leakage and reflectivity of the subcircuit. An example is shown in Fig~\ref{fig:lookup_map}. 
Non-ideal decoupling causes light leak into other parts of the circuit, which is a problem for achieving high-fidelity operations and scaling up. 
Fig~\ref{fig:leakage_analysis} shows the leakage comparison between applying 0~V and 7~V to electrode 4.
Based on our ideal simulation following the electrode configuration shown in the main text, high voltages (see Fig~\ref{fig:leakage}) are required to decrease the coupling between boundary waveguides close to 0 to minimize the leakage with our device.
In our experiments, we limit the amplitude of voltage to 10~V to prevent the chip from potential damage. Based on the simulation (see Fig~\ref{fig:leakage}), about~40\% leakages for both DCs are expected. 



To minimize light power crosstalk between subcircuits and avoid laser interference, we operated subcircuits at a distance with a number of waveguides in between.

\begin{figure}
    \centering
    \includegraphics[width=1\columnwidth]{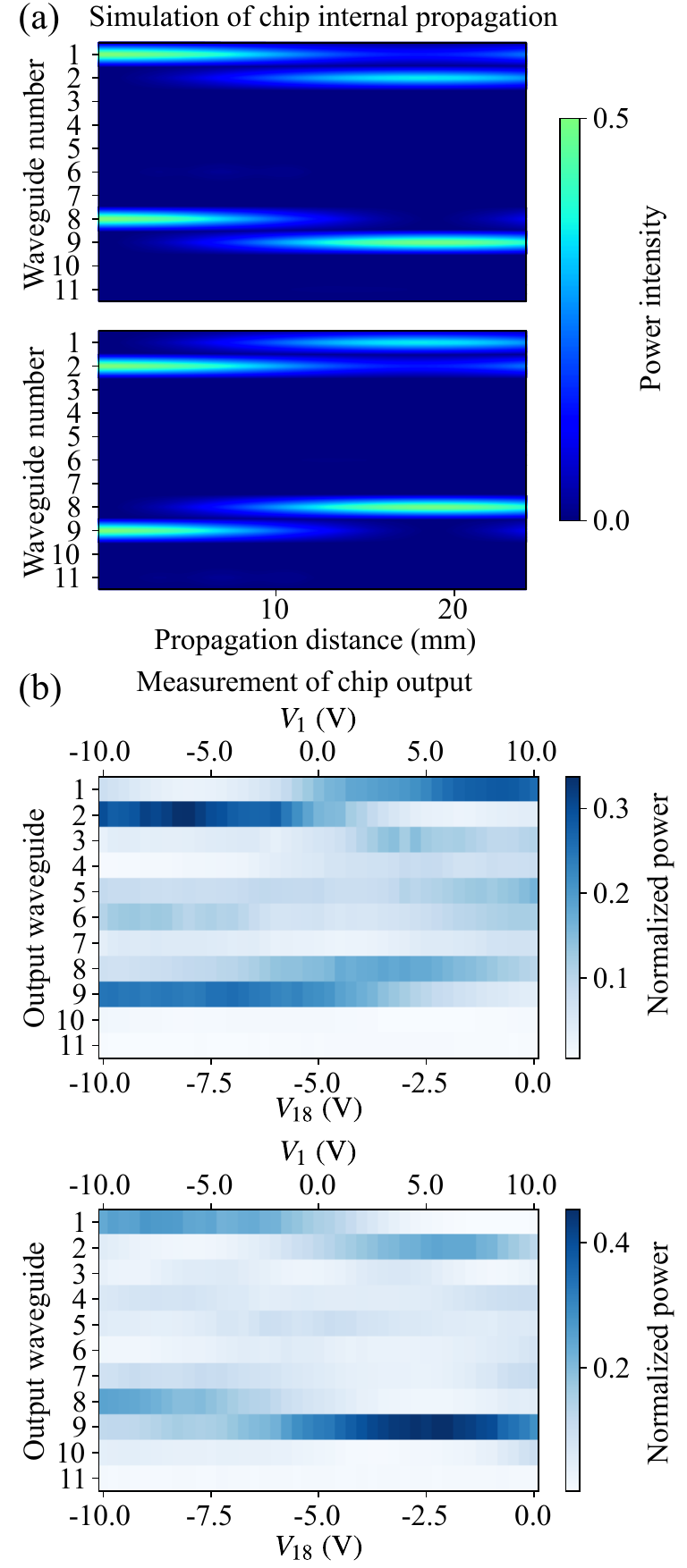}
    \caption{\textbf{Parallel control of DC$_{1,2}$ and DC$_{8,9}$.} (a) Numerical simulations of the light propagation inside the chip for input waveguides 1 and 8 (top) and 2 and 9 (bottom) respectively, where DC$_{1,2}$ and DC$_{8,9}$ are decoupled and the voltage on electrode 1 and 18 and is 0~V. (b) Measured voltage-dependent output power of the RWA for input waveguides 1 and 8 (top) and 2 and 9 (bottom) based on the implementation of the electrode configuration in the main text).}
    \label{fig:parallel_output}
\end{figure}

In Table~\ref{fig:parallel_table}, voltages used for quantum gates implementation of each subcircuit are reported and The voltages are selected based on reflectivities calculated based on the map shown in Fig~\ref{fig:parallel_output}.

\begin{table}[]
\begin{tabular}{|l|l|l|l|}
\hline
Gate (DC$_{12}$) & $V_1$ (V) & Gate (DC$_{89}$) & $V_{18}$ (V) \\ \hline
\textit{I}    & 8.0             & \textit{I}    & -2.2             \\ \hline
\textit{H}    & 0.5             & \textit{H}    & -6.5             \\ \hline
\textit{X}    & -6.4            & \textit{X}    & -9.8             \\ \hline
\end{tabular}
\caption{\textbf{Voltages used to implement parallel quantum gates in two subcircuits.} The voltages are selected from fitted reflectivity calculated based on the measured lookup map.}
\label{fig:parallel_table}
\end{table}

\begin{figure}[h!]
    \centering
    \includegraphics[width=1\columnwidth]{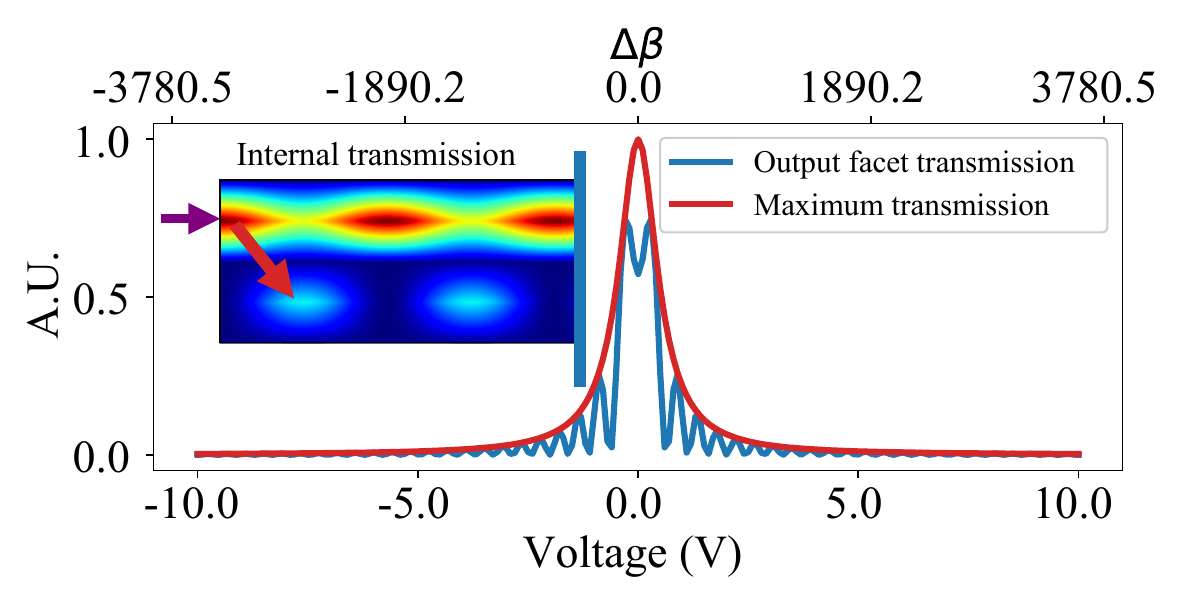}
    \caption{\textbf{The principle behind voltage-controlled reflectivity tuning.} Numerical simulations show the maximum transmission inside a coupled two-waveguide system (red) and the reflectivity at the output facet (blue) can be achieved by tuning the mode difference. The voltage is applied to one of the side electrodes to tune the propagation coefficient of one waveguide only. The inset shows the simulated internal propagation. Purple and red arrows in the inset indicate the input light and maximum transmission between waveguides respectively.}
    \label{fig:mode_tuning}
\end{figure}

\subsection*{Refectivity tuning}
A simulation of a coupled two-waveguide system is created to demonstrate the principle of reflexivity tuning in Fig~\ref{fig:mode_tuning}. 
The tunable reflectivity range of the system depends on its Hamiltonian and total coupled length. When we tune the mode of one of the waveguides in a waveguide pair, we expect the reflectivity to change at the end of the propagation in the device (see blue curve in Fig~\ref{fig:mode_tuning}).
Output facet reflectivity is defined as:
\begin{equation}
    R = 1 - T.
\end{equation}
Where T is the transmissivity of the device.
Achieving a 0-1 reflectivity requires optical mode-matching and correct propagation length~\cite{doi:https://doi.org/10.1002/0471213748.ch18}. We change the reflectivity by tuning the optical mode only, which creates mode-mismatching and results in the relatively low fidelity of the X gate implemented by DC$_{89}$. In principle, this can be compensated by tuning the coupling between waveguide pairs. Therefore, we applied an extra voltage on electrode 16. We did not fully compensate for the effect, because of the errors in electrode fabrication.

\begin{figure}[h!]
    \centering
    \captionsetup{labelfont={color=black}}
    \includegraphics[width=1\columnwidth]{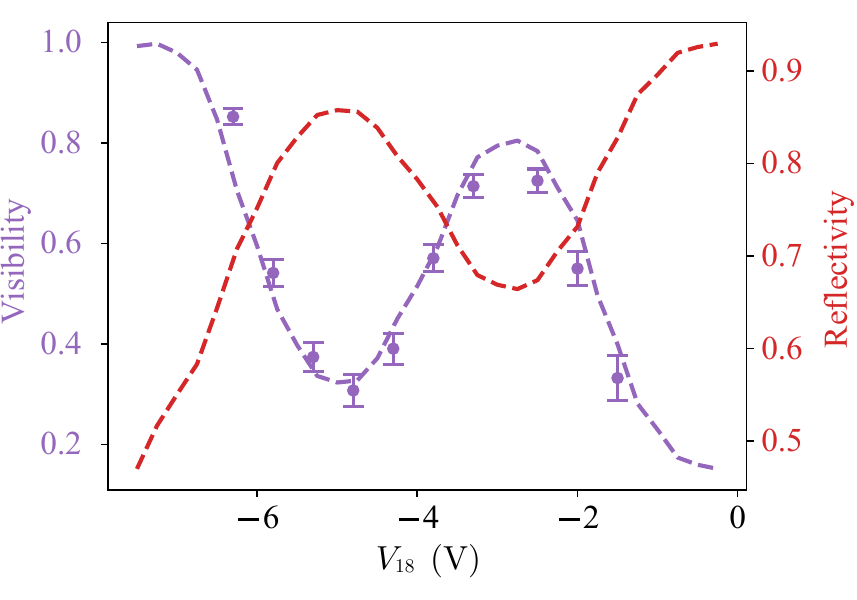}
    \caption{\textbf{The HOM interference visibility and reflectivity of subcircuit DC$_{8,9}$ as a function of $V_{18}$.} The two-photon quantum interference measurements confirmed the classical light characterization.}
    \label{fig:vis_89}
\end{figure}

\subsection*{Reflectivity and Visibility} 
In the quantum interference experiments \cite{hong_measurement_1987}, two photons generated from the SPDC source are launched into input WG$_1$ and WG$_2$ of the device. We measure the two-photon coincidence rate from output WG$_1$ and WG$_2$ while scanning through a relative arrival delay line.

The reflectivity is given by
\begin{align}
    \eta&=\frac{r}{1+r}, \\
    r& = \sqrt{\frac{P_{11}{\times}P_{22}}{P_{12}{\times}P_{21}}}
    \label{1.1.9}
\end{align}
where $P_{mn}$ is the detected power from WG$_n$ with light injecting in WG$_m$~\cite{laing2012super}.

The Hong-Ou-Mandel (HOM) two-photon coincidence measurements $ R_{\text{co}}$ are fitted by a Gaussian function with a linear term (used to compensate for the small amount of decoupling while scanning the physical delay line~\cite{laing_high-fidelity_2010}) as shown below
\begin{align}
    R_{\text{co}}=(a_0x+a_1)\left(1-a_2e^{-\frac{(x-a_3)^2}{2{a_4}^2}}\right),
    \label{1.1.g}
\end{align}
where $x$ is the scanned relative delay length between two-photon paths and $a_i$ are fitting parameters. 

Defining the full width at half maximum of the fitted Gaussian function $\alpha =2\sqrt{2 \log 2}a_4$, the maximum coincidence count can be estimated as
\begin{equation}
     N_{\text{max}}= \frac{1}{2} \left( \hat{R}_{\text{co}}(a_3 - \alpha/2) + \hat{R}_{\text{co}}(a_3 + \alpha/2) \right),
\end{equation}
while $N_\text{min}$ is the minimum co-incidence count from the measurement. 
The depth of a dip indicates the degree of quantum interference in the HOM measurement, which is given by the visibility from the fit $\bar{V}=a_2$.
The ideal visibility is given by 
\begin{equation}
    \bar{V}_{\text{ideal}}=\frac{2\eta(1-\eta)}{1-2\eta+2\eta^{2}}.
    \label{1.1.10}
\end{equation}

The visibility calculation is valid for 2$\times$2 subcircuits in a larger system~\cite{peruzzo2011multimode,laing2012super} and confirmed by the measurements of DC$_{8,9}$ shown in Fig~\ref{fig:vis_89}.

The error in the HOM dip visibility is given by:
\begin{equation}
    \epsilon_{V}=\frac{N_{\text{min}}}{N_{\text{max}}}\sqrt{\frac{1}{N_{\text{max}}}+\frac{1}{N_{\text{min}}}}.
\end{equation}

The non-unit visibility is attributed to several factors, including photon polarization and wavelength mismatches, as well as the reflectivity error caused by intra-pulse drift occurring within the non-biased square pulse modulation (see `Experimental setup' section).

\color{black}
\subsection*{Architecture loss comparison} 
A universal $N$-mode multi-port interferometer that implements the state-of-the-art Clements scheme requires $N(N-1)/2$ Mach-Zehnder interferometers (MZIs) with a maximum circuit depth of $N$. Each MZI contains 4 bending sections and experiences a constant loss of 0.2~dB, as reported in many advanced works~\cite{carolan_universal_2015, 8351983,bao2023very}. This results in a total loss of $N\times0.2$~dB. Compared to the Clements scheme, the WA-based scheme~\cite{saygin_robust_2020} has the same circuit depth $N$ but experiences lower bending losses due to the half number of bending sections the photons will experience.

RWA devices can be considered only to experience propagation loss. Many material platforms with high refractive index contrast enable the fabrication of low propagation loss waveguides, typically achieving losses less than $\sim$0.1~dB~cm$^{-1}$~\cite{politi_silica--silicon_2008,krasnokutska_ultra-low_2018,blumenthal2018silicon,grafe_integrated_2020}. 
At present, there are no studies discussing the required length for the coupled region to implement universal operations. This is an area that is currently under active research within the controllability field. However, scaling up RWA devices only requires adding more waveguides and corresponding tuners, making them promising candidates for experiencing lower photon losses in building large-scale photonic circuits.

\begin{figure}[h!]
    \centering
    \captionsetup{labelfont={color=black}}
    \includegraphics[width=1\columnwidth]{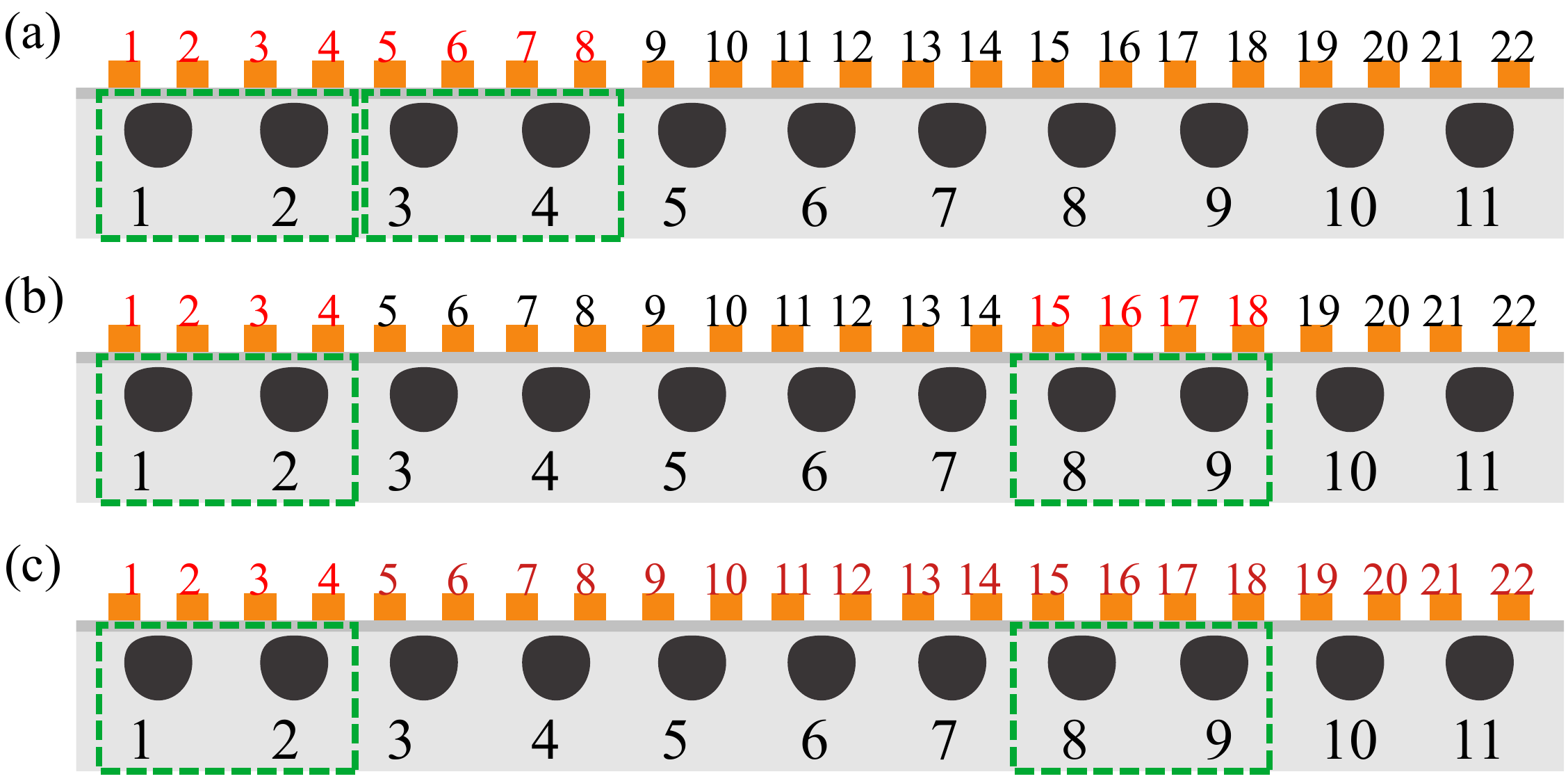}
    \caption{\textbf{Configurations for parallel $X$ gates optimization. Electrodes with red labels are used to control the subcircuits indicated with green dashed squares.} (a) Configuration 1. (b) Configuration 2. (C) Configuration 3.}
    \label{fig:optimization_configurations}
\end{figure}

\begin{figure}[h!]
    \centering
    \captionsetup{labelfont={color=black}}
    \includegraphics[width=1\columnwidth]{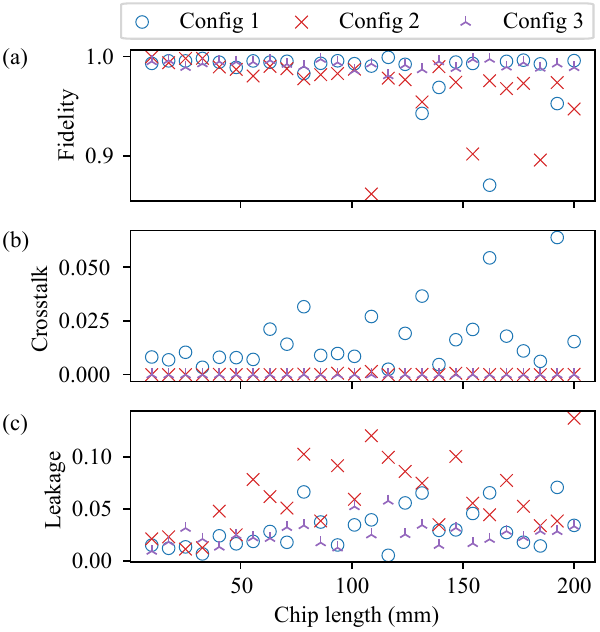}
    \caption{\textbf{Optimized parallel $X$ gates performance for three configurations averaged over the two subcircuits.} (a) Gate fidelity. (b) Crosstalk. (C) leakage.}
    \label{fig:optimization_results}
\end{figure}

\subsection*{Reducing crosstalk and leakage}
In this section, we present simulations of optimizing parallel $X$ gates using three subcircuit configurations, as illustrated in Fig~\ref{fig:optimization_configurations}. In configuration 1, electrodes 1 to 8 are used to control the adjacent subcircuits DC$_{12}$ and DC$_{34}$. In configuration 2, electrodes 1 to 4 and 15 to 18 control subcircuits DC$_{12}$ and DC$_{89}$, which are separated by 5 waveguides. In configuration 3, the electrodes used for parallel operations are extended to include all 22 electrodes.

The optimization is performed using the sequential least squares programming method~\cite{kraft1988software} with voltage amplitudes constrained to be within 10~V. 
The performance of the target parallel gates, $X$ and $X$, for two subcircuits is optimized.
The objective function to minimize is defined as follows: $(1-F_{1})^2+(1-F_{2})^2+P_{\text{crosstalk}_{1}}^{2}+P_{\text{crosstalk}_{2}}^{2}+P_{\text{leakage}_{1}}^{2}+P_{\text{leakage}_{2}}^{2}$. Here $1-F_{n}$ represents the infidelity of $X$ gate operated on subcircuit $n$, $P_{\text{crosstalk}_{n}}$ is the sum of light power crosstalk for subcircuit $n$ operations (two inputs), and $P_{\text{leakage}_{n}}$ is the sum of leakage for subcircuit $n$ operations (two inputs). The optimization process involves 100 attempts with a random initial guess of the voltages for each configuration, and the chip length varies from 10 to 200~mm. The static device has a fixed random Hamiltonian. The device model details can be found in~\cite{yang_topological_2023}.

The optimization results for parallel $X$ gates are depicted in Fig~\ref{fig:optimization_results}.
Regarding fidelity as a function of chip length, Configuration 1 outperformed Configuration 2 in some cases but performed worse in others. This suggests that the fidelity of parallel operations for adjacent or separated waveguide pairs depends on the physical parameters of the fabricated device. However, the choice of adjacent waveguide pairs can result in higher crosstalk. In addition, Configuration 3, which has more electrodes, consistently performed better, especially in terms of reducing leakage, due to its enhanced control capabilities. 

In this optimization process, each iteration requires four measurements. While these optimization results are promising, implementing such an online optimization scheme experimentally is very challenging, as experimental noise can easily lead the optimization in the wrong direction. Alternatively, machine-learning methods such as the graybox approach demonstrated in our previous work on a small-scale system~\cite{youssry_experimental_2022} might offer a more feasible solution.